**Topological descriptor of thermal conductivity in amorphous materials**

Emi Minamitani[1,2,3*], Takuma Shiga[4,3], Makoto Kashiwagi[5,3], Ippei Obayashi[6,3]

1. Institute for Molecular Science, Okazaki 444-8585, Japan
2. The Graduate University for Advanced Studies, Okazaki 444-8585, Japan
3. JST, PRESTO, Kawaguchi, Saitama, 332-0012, Japan
4. Department of Mechanical Engineering, The University of Tokyo, Tokyo 113-8656, Japan
5. Graduate School of Science and Engineering, Aoyama Gakuin University, Sagamihara 252-5258, Japan
6. Cyber-Physical Engineering Informatics Research Core (Cypher), Okayama University, Okayama 700-8530, Japan

* Corresponding author: eminamitani@ims.ac.jp

**Abstract**

**Quantifying the correlation between the complex structures of amorphous materials and their physical properties has been a longstanding problem in materials science. In amorphous Si, a representative covalent amorphous solid, the presence of a medium-range order (MRO) has been intensively discussed. However, the specific atomic arrangement corresponding to the MRO and its relationship with physical properties, such as thermal conductivity, remains elusive. We solved this problem by combining topological data analysis, machine learning, and molecular dynamics simulations. Using persistent homology, we constructed a topological descriptor that can predict thermal conductivity. Moreover, from the inverse analysis of the descriptor, we determined the typical ring features correlated with both the thermal conductivity and MRO. The results provide an avenue for controlling material characteristics through the topology of the nanostructures.**

## I. Introduction

The structure of amorphous materials is characterized by the absence of long-range order (LRO) and the presence of some medium-range order (MRO)[1] beyond the short-range order (SRO). Revealing the quantitative correlation between the structural and physical properties of amorphous materials remains a challenging task. Thermal conductivity is a fundamental physical property that shows unique behavior in amorphous materials owing to the strong interaction between lattice vibrations and disorders. The lack of LRO reduces the lattice thermal conductivity by several orders than that of a crystal with the same stoichiometry[2]. The heat carriers, vibrational modes, in amorphous materials are generally classified into propagating and non-propagating modes[3,4]. The former is exhibited in the low-frequency range and has characteristics similar to those of phonons in the crystal. In contrast, the latter carries heat in a diffusive manner rather than propagating energy as phonons do in the crystal. It is expected that the MRO affects the propagation and diffusion of these vibrational modes, and thus the thermal conductivity.

The structural signature of the MRO and thermal conductivity have been intensively investigated in amorphous Si, which is widely used in various applications, from solar cells to image sensors[5,6]. The presence of MRO has been investigated theoretically through the dihedral angle distribution with ring statistics[7–11] and detected using fluctuation electron microscopy[12,13]. The lattice thermal conductivity has been measured or simulated in samples with a wide variation in preparation methods and sizes[14–20]. Recently, it has been proposed that the MRO affects the thermal conductivity mediated by the propagating modes[21].

Previous studies indicated that determination of the atomic structure corresponding to the MRO and extraction of the correlation between the MRO and the lattice thermal conductivity are essential to precisely control the thermal properties of amorphous Si. However, these tasks remain challenging because determining the essential features of MRO from traditional structural analysis is difficult, such as the pair distribution function and bond-orientation order analysis.

Recently, persistent homology[22–25], an emerging technique in the field of topological data analysis[26,27], has been employed to describe the atomic structures corresponding to MRO in $SiO_2$ glass[28–31], metallic glass[28,32,33], and amorphous ice[34]. The advantage of persistent homology is that multiscale topological information can be extracted from complicated structures[35]. For the analysis of persistent homology, we considered a growing sequence of network structures for given data points with different scale lengths defined by the filtration procedure. A schematic of the filtration procedure for a two-dimensional case is shown in Fig. 1. As can be seen, we considered circles centered at the respective data points. Subsequently, the radius of each circle gradually increases. The sequence of increase in radius is often referred to as "time." At some radius, the circles start to intersect with each other, and we set an edge between the centers of the circles. When the edges form a closed ring, this corresponds to a topological feature called a "cycle." As the radius further increases, the ring gets fully covered by circles. This is interpreted as the cycle converting into another class of topological features called a "boundary." The topology of the data is represented by the pairs of birth and death times at which the cycle appears and is converted into a boundary. This filtration procedure can be generalized into three or higher dimensions using spheres or hyperspheres. The two-dimensional visualization of birth and death time pairs is called a persistence diagram (PD). The inverse analysis technique[36] provides a characteristic structure that generates a specific birth–death pair in the PD. Moreover, the vectorization of a PD[37,38] enables the creation of a descriptor used in the machine learning procedure.

Several recent studies have extracted unique geometric characteristics by combining the analysis of PDs and machine learning[33,39,40]. These methods are expected to be powerful techniques for revealing the correlation between the structural and physical properties of amorphous materials.

In this study, using persistent homology, we constructed reliable descriptors for lattice thermal conductivity, reflecting the topological features of the MRO in amorphous Si. A structural model of amorphous Si was generated via the melt-quench method using classical molecular dynamics (MD), where the system temperature was increased above the melting temperature and then gradually cooled to room temperature. The difference in structural characteristics was introduced by changing the cooling rate in the MD simulation from $10^{14}$ to $10^{11}$ K/s. We selected 570 snapshots from the equilibrated MD simulation after the melt-quench procedure, and the thermal conductivity mediated by non-propagating modes and the PD were evaluated for each structure. Next, we constructed a descriptor of the topological features using the persistent image of the PD. We demonstrated that supervised training for the dataset of these descriptors and lattice thermal conductivities could achieve accurate predictions. In addition, from the inverse analysis by volume-optimal cycle, we determined the typical ring features correlated with the thermal conductivity and MRO. Our study is the first demonstration of the prediction of the physical properties of amorphous materials based on topological features. In addition, our results illustrate the hidden relationship between MRO and the physical properties of amorphous materials. This study opens an avenue for controlling material characteristics through the topology of nanostructures.

## II. Methods

### A. Molecular dynamics simulation

The amorphous model structures in this study were generated using a melt-quench algorithm and classical MD simulations with the LAMMPS package[41]. The simulation unit cell size was 32.76 Å × 32.76 Å × 32.76 Å with periodic boundary conditions, and 1,728 Si atoms were included in the unit cell. We employed the Stillinger–Weber interatomic potential[42]. All MD simulations were performed using an NVT ensemble. The system temperature was controlled using a Nosé–Hoover thermostat[43,44], with 1 fs time step. The initial crystal structure was first melted by increasing the temperature from 300 to 3,300 K at the rate of $3.0 \times 10^{12}$ K/s. After the equilibration step of 500 ps, the systems were cooled to 300 K at different cooling rates ranging from $1.0 \times 10^{14}$ to $1.0 \times 10^{11}$ K/s. The systems were maintained at 300 K for 500 ps to stabilize, and then the amorphous model structures were sampled from the trajectory of the equilibrium MD steps. We sampled 30 structures at 19 different cooling rates. The atomic coordinates of the amorphous structures were visualized using VESTA[45,46].

### B. Thermal conductivity calculation

Vibrational modes in amorphous solids can be classified into propagons, diffuson, and locons. Propagons are low-frequency modes that retain elastic wave characteristics similar to acoustic phonons. Diffusons are delocalized vibrations but have no apparent periodicity in their eigenvector. Locons are strongly localized modes. Heat transport in amorphous solids is governed by propagons (propagating modes) and diffusons (non-propagating modes). The thermal conductivity by the latter cannot be described by the phonon gas model combined with the Boltzmann transport equation, which is widely used in crystals. The Allen–Feldman theory[3,4] is one of the alternative methods to compute diffuson thermal conductivity.

The starting point of the Allen-Feldman theory is the following definition of thermal conductivity based on the Kubo formula.

$$\kappa_\mu = \frac{V}{T}\int_0^\beta d\lambda \int_0^\infty dt\, \langle S_\mu(-i\hbar\lambda) S_\mu(t)\rangle\ . \quad (1)$$

In the above, $\mu = x, y, z$; $V$ is the system's volume; $T$ is the temperature and $\beta = 1/k_B T$; $S_\mu$ is the $\mu$ component of the heat flux operator $\boldsymbol{S}$. In the harmonic approximation, the heat flux operator can be decomposed into the contributions of respective modes. The decomposed heat flux operator $\boldsymbol{S}_{mn}$ is defined by the eigenvectors of vibrational modes and the dynamical matrix $D_{\alpha\beta}^{ij}$, as follows:

$$\boldsymbol{S} = \sum_{mn} \boldsymbol{S}_{mn} a_m^\dagger a_n\ , \qquad (2)$$

$$\boldsymbol{S}_{mn} = \frac{\hbar}{2V}\boldsymbol{V}_{mn}(\omega_m + \omega_n)\ , \quad (3)$$

$$\boldsymbol{V}_{mn} = \frac{i}{2\sqrt{\omega_m\omega_n}}\sum_{\alpha\beta}\sum_{\gamma\gamma'}(\boldsymbol{R}_\gamma - \boldsymbol{R}_{\gamma'})D_{\alpha\beta}^{\gamma\gamma'} e^\alpha(\gamma, m)e^\beta(\gamma', n)\ . \quad (4)$$

In the above, $m, n$ are the indices of the vibrational modes; $a_m^\dagger$ $(a_n)$ corresponds to the generation (annihilation) operator of a vibrational mode; $\alpha, \beta = x, y, z$; $\omega_m$ is the vibrational frequency of the mode $m$; $\gamma, \gamma'$ are indices that distinguish the atoms in a supercell; $R_\gamma$ is the position vector of atom $\gamma$; $e^\alpha(\gamma, m)$ is the polarization of the eigenvector of the vibrational mode $m$ at atom $\gamma$ along $\alpha$ direction. Note that the definition of the dynamical matrix $D_{\alpha\beta}^{\gamma\gamma'}$ is

$$D_{\alpha\beta}^{\gamma\gamma'} = \frac{1}{\sqrt{m_\gamma m_{\gamma'}}} \frac{\partial^2 E}{\partial u_\gamma^\alpha \partial u_{\gamma'}^\beta}, \quad (5)$$

where $E$ is the total energy and $u_\gamma^\alpha$ is the displacement of the atom.

The average of thermal conductivity along $\mu = x, y, z$ directions is often expressed by

$$\kappa = \frac{1}{V} \sum_m C_m(T) D_m, \quad (6)$$

where $C_m(T) = \frac{\hbar\omega_m^2}{T}\left[-\frac{\partial n_m}{\partial \omega_m}\right]$ is the specific heat of mode $m$ and $D_m$ is the mode diffusivity defined by

$$D_m = \frac{\pi V^2}{3\hbar^2 \omega_m^2} \sum_n |\boldsymbol{S}_{mn}|^2 \, \delta(\omega_m - \omega_n). \quad (7)$$

In this study, we evaluated $D_m$ and $\kappa$ using GULP package[47]. The structures were optimized before calculating the thermal conductivity to eliminate the appearance of imaginary frequencies. During the calculation of diffusivities, we approximated the delta function by the Lorentzian with a width of five times the average mode frequency spacing (~ 5 × 0.124 cm$^{-1}$). Furthermore, to avoid ambiguities in the definition of the threshold frequency of the propagating modes[14,48–50], we evaluated the contribution from all vibrational modes by diffuson diffusivity[3]. We note that this approximation caused underestimation of the contribution from the propagating mode. This underestimation did not affect our conclusion on the relationship between local coordinates and thermal conductivities, as we discussed in Sec. IIIC and Supplementary Materials.

## C. Persistent homology analysis

The HomCloud code[51] was used to analyze the persistent homology of amorphous models. The PDs were evaluated for the optimized structures used in the thermal conductivity calculations. We focused on the first homology. For machine learning analysis, the diagrams were converted into vectors using the persistent image method[37,38]. We chose a linear weighting for the distance from the diagonal line in the construction of persistent image vectors. The standard deviation of the Gaussian used in the persistent image method was set to 0.06. We focused on the square region $[0.0, 6.0]_{\mathrm{birth}} \times [0.0, 6.0]_{\mathrm{death}}$, which was divided into 300 × 300 meshes.

### D. Machine learning

Ridge regression and principal component analysis (PCA) using persistent image vectors were conducted using the Scikit-learn package[52]. For ridge regression, 80% of the 570 samples were used as training data and the rest as test data. The regularization parameter was set to 1.0, which resulted in the average $R^2$ score of 0.992 for a five-fold cross-validation. In PCA, 40 samples were used, among which, 20 samples were taken from the models generated at the cooling rate of $1.0 \times 10^{14}$, and the rest were taken from those at $1.0 \times 10^{11}$ K/s.

## III. Results and Discussion

### A. Regression model of thermal conductivity using persistent homology

To quantitatively investigate the relationship between PD and thermal conductivity, we constructed a regression model of thermal conductivity based on persistent homology. First, we generated samples with various degrees of randomness in the structures through melt-quench simulations with different cooling rates[53–55]. Fig. 2 (a) shows the correlation between the cooling rate and thermal conductivity. A slower cooling rate generally results in a higher thermal conductivity compared with a faster cooling rate. Figs. 2 (b) and (c) present the averaged PDs obtained from 30 structures generated at cooling rates of $10^{11}$ and $10^{14}$ K/s, respectively. The difference appears in the $[1.4,1.7]_{\mathrm{birth}}$ region. In Fig. 2 (b), a sharp vertical line appears in $[1.4,1.5]_{\mathrm{birth}} \times [3.5, 5.0]_{\mathrm{death}}$, which fades and shifts toward a larger birth–smaller death zone in (c). The birth time (~1.5 Å$^2$) of the sharp vertical line in (b) is explained by the squared value of half of the nearest neighbor distance in amorphous Si (~2.5 Å). Even in the amorphous phase, the variation in the nearest neighbor distance is within 0.3 Å; thus, the distribution of the birth becomes narrow. The death time corresponds to the maximum radius of the ball that can pass through the cycle. Therefore, the death time reflects both the size and shape of the cycle. As shown later, the cycles can have up to 44 vertices, which indicates a wide range of sizes. Furthermore, cycles with the same number of vertices and bond lengths have the same birth times but can have different death times when they have different combinations of interior angles. This results in a wide range of death times to form a vertical line. The differences between (b) and (c) are discussed in the following sections.

Further we performed a regression analysis between the persistent image vector generated from the PD and thermal conductivity. Fig. 3 (a) shows the results of the fitting using ridge regression. The root mean squared error of the prediction was 0.007 W/mK, indicating a strong correlation between persistent homology and thermal conductivity. Fig. 3 (b) shows the reconstructed PD from the ridge coefficient of the fitted model. The color level indicates positive and negative influences on the thermal conductivity. The presence of birth–death pairs in the blue region corresponds to a higher thermal conductivity, and that in the red region corresponds to a lower thermal conductivity. The distribution of the area with dark blue and red colors matches the area where a difference is observed in the PDs at cooling rates of $10^{11}$ and $10^{14}$ K/s.

### B. Relationship between local structure, persistent homology, and thermal conductivity

To clarify the atomic configurations that determine the dependence of thermal conductivity on the cooling rate, we focused on two extreme cases. For the models with a higher thermal conductivity, we selected 20 structures generated at the slowest cooling rate of $10^{11}$ K/s. The thermal conductivities of these samples were in the range of 0.997–1.017 W/mK. For the models with a lower thermal conductivity, we selected the same number of structures generated at the fastest cooling rate of $10^{14}$ K/s, where the thermal conductivities were 0.776–0.783 W/mK. Hereafter, we define the former group as the slowest-cooled model and the latter as the fastest-cooled model. We performed a PCA for the persistent image vectors obtained in these models. The results of the PCA on 40 models, 20 of which were slowest-cooled and the rest were fastest-cooled, are shown in Fig. 4. From Fig. 4 (a), it can be observed that the first principal component properly separates the slowest- and fastest-cooled models. Therefore, we reconstructed a PD from the coefficient of the first principal component, as displayed in Fig. 4 (b).

In Fig. 4 (b), the coefficient for each element in the persistent image vector is projected onto the mesh of the square region $[0.0, 6.0]_{\mathrm{birth}} \times [0.0, 6.0]_{\mathrm{death}}$. As shown in Fig. 4 (a), the slowest- (fastest-) cooled models have negative (positive) first principal components. Considering that the elements of the persistent image vector are non-negative, the blue and red areas correspond to the birth–death pair position observed in the slowest- and fastest-cooled models, respectively. The overall features of the PD reconstructed from the coefficient of the ridge regression and the first principal component were consistent. The results in Fig. 4 (b) are summarized as follows:

1. Pairs in $[1.38\text{-}1.52]_{\mathrm{birth}} \times [3.52\text{-}4.90]_{\mathrm{death}}$ for the slowest-cooled models have positive contributions to the thermal conductivity.
2. Pairs in $[1.52\text{-}1.86]_{\mathrm{birth}} \times [2.86\text{-}4.11]_{\mathrm{death}}$ for the fastest-cooled models have negative contributions to the thermal conductivity.

These two characteristics can be partly understood by the difference in the randomness of the bond lengths in the slowest- and fastest-cooled models. The narrow birth time range in (1.) directly corresponds to the sharp distribution of the bond length in the slowest-cooled models. Considering that the longer death time corresponds to the cycles with a large size, the wider distribution of the death time within a narrow birth time reflects the uniformity of the bond length. In contrast, the larger inhomogeneity of the bond length in the fastest-cooled samples tends to shift the distribution of the birth-death pair to the larger birth time region because the birth time is determined by the maximum bond length in the cycle. Although there are cycles with large sizes in the fastest-cooled samples, the wider variety of the birth time blurs the distribution of the pairs with a large death time. Thus, the PCA coefficients of the fastest-cooled samples are mainly distributed in the $[1.52\text{-}1.86]_{\text{birth}} \times [2.86\text{-}4.11]_{\text{death}}$ region.

To discuss the difference quantitatively, we performed an inverse analysis to obtain the atomic configuration for the dominant birth–death pairs with large PCA coefficients. We extracted the birth–death pairs in the slowest-cooled (fastest-cooled) models by filtering with the condition that the coefficient of the first principal component was $< -0.06$ ($> 0.02$). To realize similar length scales of the pairs extracted from the two cases, we clipped the death time within [3.7, 3.9]. Using these procedures, we selected 785 (1,266) pairs in the slowest-cooled (fastest-cooled) models. By tracing the filtration procedure, we could determine the candidates of the cycle that correspond to each birth-death pair. Sometimes, several candidates have equivalent topological characteristics. There are several concepts for the choice of the representative cycle in such a case. Here, we adopted the volume-optimal cycle[36], in which, we chose a cycle that minimizes the internal volume. As we focused on the first homology, the shape of a volume-optimal cycle becomes a ring structure. The number of vertices (boundary points) in each volume-optimal cycle is presented in Fig. 5. The results show that ring structures with five vertices are the major volume-optimal cycles in both the slowest- and fastest-cooled models. We also analyzed the number of children's birth–death pairs for each volume-optimal cycle, as shown in the lower half of Fig. 5. The children of a parental cycle correspond to the approximate partial structures generated between the birth and death of the parental cycle, which reflects the hierarchy in the topological features. We observed that the five-vertex cycles do not have children's birth–death pair; thus, they can be considered as the primary units governing the thermal conductivity.

Moreover, we conducted a traditional structural analysis to investigate the characteristics of these five-vertex cycles. Fig. 6 exhibits a histogram of the lengths of the edges and angles in the five-vertex cycles, together with a comparison between the PCA for these data and PD. Figs. 6 (a) and (b) show that the randomness in the edge lengths and angles is larger in the fastest-cooled models than in the slowest-cooled models. Thus, we attempted to classify the slowest- or fastest-cooled model via PCA using the five values of edge length and angle in each structure. As shown in Figs. 6 (c) and (d), the PCA using the values of edge length exhibited a better separation than that using the angle. This indicates that the strength of the randomness of the edge length is the major difference between the five-vertex cycles in the slowest- and fastest-cooled models. Considering that the edge length and bond angle are typical descriptors for the SRO, the insufficient separation of the two models with these values would indicate the length scale of the five-vertex cycles beyond the SRO. The successful separation of the two models by PD shown in Fig. 6 (e) can be attributed to the multiscale character of the persistent homology analysis.

In addition to the evaluation of the primal five-vertex cycles, we analyzed the volume-optimal cycles for children's birth–death pairs. Fig. 7 (a) presents the distribution of the number of vertices in each child's volume-optimal cycle. From these results, we focused on the children's cycles with three, four, five, and six vertices. The distributions of each type of children's cycles, together with the PD, are provided in Fig. 7 (b) and (c). A notable difference appears in the four-vertex children's cycles between the slowest- and fastest-cooled models. In the slowest-cooled models, the number of four-vertex children's cycles was much smaller than that in the fastest-cooled models. In addition, the distribution of the four-vertex children's cycles in $[1.6,2.5]_{\mathrm{birth}} \times [2.5,3.0]_{\mathrm{death}}$ matches the red area appearing in the reconstructed PD from the coefficients of ridge regression and the first principal component shown in Figs. 3 (b) and 4 (b). Considering that the birth times of these four-vertex children's cycles are greater than those of most of the primal five-vertex cycles, these structures are considered to be the representative perturbations that make the topology in the slowest- and fastest-cooled models different in the scale larger than the primal five-vertex cycles.

Next, we discuss the relationship between our results and MRO in amorphous Si. As discussed above, the five-vertex cycles without children strongly correlate with thermal conductivity. This indicates that these five-vertex cycles are the minimal components of the MRO that affect the thermal conductivity of amorphous Si. In addition, the hierarchy of the children's cycles indicates a strong difference in structures with high and low thermal conductivity. The four-vertex children's cycles observed in the low-thermal-conductivity models would correspond to the perturbation that breaks the MRO.

The correlation between the MRO and the thermal conductivity is further supported by comparing the radial distribution function (RDF) obtained from the entire amorphous structure and the partial region limited to the volume-optimal cycles and their children's cycles correlating with the thermal conductivity. As shown in Fig. 8 (a), the RDF also depends on the cooling rate. At each cooling rate, there is a sharp first peak at ~ 2.5 Å and a second peak at ~ 3.8 Å. The second peak is accompanied by a shallow peak around 3.3 Å. The first peak corresponds to the SRO, while the second peak corresponds to the MRO with the shortest length scale[1]. We employed a ridge regression model to predict the thermal conductivity from the RDF. For the vector data used in the ridge regression, we used the RDF obtained by binning the pair-wise distance in the range 2.0–6.0 Å into 200 bins. As presented in Figs. 8 (b) and (d), a smaller L2 regularization parameter ($\alpha$) indicates a better prediction. However, the distribution of the coefficient is clearer with a large regularization parameter; thus, we show the ridge coefficients obtained with $\alpha = 50.0$ in Fig. 8 (c). A comparison between Figs. 8 (a) and (c) indicates that the sharper first and second peaks increase the thermal conductivity, whereas the tail of the first peak and the shallow peak around 3.3 Å are associated with decreased thermal conductivity.

Next, we evaluated the RDF using partial structures in the slowest- and fastest-cooled models. The RDF restricted to the volume-optimal cycles that correlate with the large positive and negative first principal components in Fig. 4 shows the enhanced difference in the shapes of the first and second peaks, as depicted in Fig. 9 (a). When we further restricted the RDF calculation to the volume-optimal cycles with five vertices, only the two peak structures associated with SRO and MRO appear, and no intensity between 2.5–3.5 Å, as in Fig. 9 (b). This indicates that these five-vertex cycles do not contain the configurations that break the MRO and correlate with both SRO and MRO. These results strongly support the idea that these five-vertex cycles are the units of the MRO. Furthermore, the four-vertex children's cycles that appear in the fastest-cooled models demonstrate a shallow peak structure at 3.3 Å in the RDF, as shown in Fig. 9 (c). This also suggests that the MRO is disturbed by the presence of four-vertex children's cycles in the fast-cooling models.

Finally, we examined how the five- and four-vertex cycles directly influence the thermal conductivity. Fig. 10 (a) shows that the overall frequency dependences of the mode diffusivities for the slowest- and fastest-cooled models are similar. However, the difference in the magnitudes of the mode diffusivities in the range 180–400 $cm^{-1}$ gives rise to the difference in the thermal conductivities of the two models, which is apparent in the cumulative thermal conductivity (Fig. 10 (b)). From the inversion participation ratio (IPR) shown in Fig. 10 (c), the mobility edge, which characterizes the transition from diffusive modes to strongly localized modes, is estimated to be ~ 570 $cm^{-1}$. Therefore, the modes in the range 180–400 $cm^{-1}$ were assigned to the diffusive modes.

The vibrational density of states (DOS) for the two models presented in Fig. 10 (d) have two typical peaks with frequencies of ~180 and ~550 $cm^{-1}$. These spectral shapes are consistent with those of previous studies on amorphous Si[10,56]. Despite the apparent difference exhibited in the second peak, the total DOS at around 180 $cm^{-1}$, which predominantly contributes to the thermal conductivity, is similar for the two models. To obtain further insight, we decomposed the DOS into the respective atoms (Fig. 10 (e)). The DOS projected onto the atoms involved in the five-vertex cycles for the slowest-cooled samples holds two-peak structures, indicating that the five-vertex cycles determine the overall frequency dependence of the total DOS. In contrast, the projected DOS related to the four-vertex children's cycles for the fastest-cooled models has a decreased second peak and a small new peak near the maximum frequency, marked by the black arrow.

From the IPR results (Fig. 10 (c)), the new peak corresponds to the strongly localized vibrational modes on the atoms in four-vertex children's cycles, and its emergence is attributed to the large deformation of the bonds yielded by the formation of the four-vertex children's cycles. This deformation results in inhomogeneity of the force constants. As shown in Eq. (4), the overlap between the eigenvectors, $e^{\alpha}(\gamma, m)e^{\beta}(\gamma', n)$, governs the diffusivity. The inhomogeneity of the force constant introduces the randomness of the atomic displacements, which reduces the overlap and hinders the diffusivity of the modes. This explains the deviation of the thermal conductivities for the two models, despite the similarity in the total DOSs.

To confirm the difference in the inhomogeneity of the force constants, we plotted a histogram of the absolute values of the dynamical matrix element $D_{\alpha\beta}^{\gamma\gamma'}$ defined as Eq. (5). As we discussed above, five-vertex cycles (four-vertex children's cycles) are the primal structure (perturbation) of MRO. Thus, we focused on the matrix elements correlated to these cycles. In Fig. 10 (f), histograms for the three sets are plotted: set 1 is $\{\left|D_{\alpha\beta}^{\gamma\gamma'}\right|$ : $\gamma$ or $\gamma$' $\in$ five-vertex cycles in slowest-cooled samples$\}$, set 2 is $\{\left|D_{\alpha\beta}^{\gamma\gamma'}\right|$ : $\gamma$ or $\gamma$' $\in$ five-vertex cycles in fastest-cooled samples$\}$, and set 3 is $\{\left|D_{\alpha\beta}^{\gamma\gamma'}\right|$ : $\gamma$ or $\gamma$' $\in$ four-vertex children's cycles in fastest-cooled samples$\}$. In every histograms, there is a peak structure in the range 0.4–0.7 eV/amu·Å$^2$. The peak in set 2 is much broader than that in set 1. And, the peak in set 3 is broader than that in set 3. These are interpreted as the force constant in the fastest-cooled samples has larger inhomogeneity than that in the slowest-cooled samples. More specifically, the MRO and its breaking correlate with the inhomogeneity of the force constants. This provides a direct connection between the local structures, MRO, and thermal conductivity of amorphous Si.

## C. Analysis in larger system size

The above results show that persistent homology links disorders in the local structure and thermal conductivity. However, the sample size used in the previous sections was relatively small. In addition, in the Allen-Feldman theory, we treated the contributions from all modes by diffuson diffusivity. Therefore, the thermal transport by low-frequency propagating modes may not have been appropriately evaluated. To rule out the possibility that the obtained results are artifacts because of the sample size or underestimation of the contribution from the propagating mode, we further analyzed a larger system containing 4,096 Si atoms with the cell size of 4.378 nm $\times$ 4.378 nm $\times$ 4.378 nm.

We focused on cooling rates of $10^{11}$ and $10^{14}$ K/s as representatives for the slow and fast cooling conditions. Ten samples of amorphous structures were prepared using these cooling rates, followed by thermal conductivity analysis using the Green-Kubo method and persistent homology analysis. Details of the calculation and results are summarized in Supplementary Materials. We evaluated the dynamic structure factor using the following definition to confirm the presence of the propagating mode in this sample size[14,18]:

$$S_{L,T}(\boldsymbol{q},\omega) = \sum_{\nu}\sum_{i}\left|u_i^{L,T} e^{i\boldsymbol{q}\cdot\boldsymbol{r}_i}\right|^2 \delta\big(\omega - \omega(q = 0,\nu)\big), \quad (8)$$

where $\boldsymbol{q}$ is the phonon wavevector, $\omega$ is the frequency, $\nu$ is the index of the modes at the gamma point, and $i$ indicates the index of the atom. $u_i^L = \hat{\boldsymbol{q}} \cdot \boldsymbol{e}(\nu, i)$ and $u_i^T = \hat{\boldsymbol{q}} \times \boldsymbol{e}(\nu, i)$, where $\hat{\boldsymbol{q}}$ is the unit vector parallel to $\boldsymbol{q}$ and $\boldsymbol{e}(\nu, i)$ is the eigenvector of the mode. As shown in Fig. 11, dispersion relationships similar to longitudinal and transverse acoustic phonons were obtained. The line width of $S_{L,T}$ is a few $cm^{-1}$ for the modes with frequencies below 100 $cm^{-1}$; such modes are considered propagating modes.

The mean value of the Green-Kubo thermal conductivity is 1.56 W/mK for the sample prepared at $10^{11}$ K/s, which is consistent with the results of a previous study. Meanwhile, the mean value for the sample prepared at $10^{14}$ K/s was 1.34 W/mK. We also evaluated the thermal conductivity using the Allen-Feldman theory for the same samples. The mean values of 1.10 and 0.81 W/mK were obtained for the samples generated at cooling rates of $10^{11}$ and $10^{14}$ K/s, respectively. The smaller value realized using the Allen-Feldman theory compared with that obtained in the Green-Kubo method is due to underestimation of the contribution from the propagating mode below 100 $cm^{-1}$, as discussed above. Nonetheless, the difference in the thermal conductivity in the slow- and fast-cooled samples is common in these two methods, which supports that the dependence of the thermal conductivity on the cooling rate is not an artifact.

As summarized in Supplementary Materials, the characteristics of the persistent homology for the 4,096-atom systems are similar to the results for the 1,728-atom systems, including the persistence diagrams and the birth-death pair distributions that characterize the structures generated at cooling rates of $10^{11}$ and $10^{14}$ K/s. Similar to the 1,728-atom systems, the five-vertex cycles are the most majority, and a notable difference appears in the number and distribution of the four-vertex children's cycle. Moreover, these characteristic structures are related to the inhomogeneity of the force constants, as in the 1,728-atom systems. These results confirm that the correlation between disorders in local structures, thermal conductivity, and persistent homology holds regardless of the sample size.

## IV. Conclusion

We demonstrated that the thermal conductivity of amorphous Si can be predicted using topological data analysis. We observed that the PD of the first homology depends on the cooling rate in the melt-quench simulation to prepare an amorphous structure. The persistent image vectors were strongly correlated with the thermal conductivity. This enabled us to construct a ridge regression model with high precision for the prediction of thermal conductivity from structural information. We also determined the unit of the MRO that governs the thermal conductivity from the inverse analysis using the volume-optimal cycle. We observed that the primary structure of MRO in amorphous Si was five-vertex cycles. Analysis of the children's cycles revealed that the MRO was perturbed by the characteristic four-vertex children's cycle in the fastest-cooling conditions, which resulted in a lower thermal conductivity. In addition, the spectral shape of the RDF derived from the generator (five-vertex cycles)/inhibitor (four-vertex children's cycles) of the MRO was clarified. Finally, we visualized the direct connection between these dominant atomic arrangements and thermal conductivity. Our results clarify the relationship between the actual local structures, MRO, and thermal conductivity in amorphous materials for the first time. This indicates further possibilities for investigating the novel relationship between complex structures and physical properties by combining topological data analysis, machine learning techniques, and computational materials science.

## Acknowledgments

This study was supported by JST, PRESTO Grant Numbers JPMJPR17I7, JPMJPR17I5, JPMJPR19I4, JPMJPR1923, JPMJPR2198, and MEXT KAKENHI 21H01816, 19H00834, 20H05884, Japan. Calculations were performed using a computer facility at the Research Center for Computational Science (Okazaki, Japan).

Figure captions

**Fig. 1. Schematic of the filtration procedure used to obtain a persistence diagram (PD) from data points.**

Schematic of the filtration procedure used to obtain a PD from the data points. Here, the network structures defined at the birth times of the two representative cycles are depicted. The red-filled polygon indicates that the ring structure formed by the edges of the polygon is converted into a boundary from a cycle.

**Fig. 2. Thermal conductivities and PDs in amorphous models.**

a) Thermal conductivity of the amorphous models generated at different cooling rates.
b) Averaged PD for 30 amorphous structures generated at the slowest cooling rate of $10^{11}$ K/s. The histogram of the birth–death pairs in the $[0,8]_{\mathrm{birth}} \times [0,8]_{\mathrm{death}}$ area was evaluated using a 128 × 128 mesh. The color bar indicates the number of birth–death pairs. We used the squared values of the radii of the spheres used in the filtration procedure as the birth and death times. Therefore, their units are $Å^2$.
c) Averaged PD for 30 amorphous structures generated at the fastest cooling rate of $10^{14}$ K/s. The calculation method for the histogram was the same as that for (b).

**Fig. 3. Ridge regression analysis using PDs.**

a) Results of prediction of thermal conductivity for test data using the trained ridge regression model
b) Reconstructed PD from the coefficient in the ridge regression model.

**Fig. 4. PCA for PDs obtained from slowest- and fastest-cooled models.**

a) First and second principal components obtained from the PCA for the fastest- and slowest-cooled models.
b) Reconstructed PD from the coefficient of PCA.

**Fig. 5. Histogram of the number of vertices for volume-optimal cycles.**

Upper: histogram of the number of vertices (boundary points) for the volume-optimal cycles attributed to the birth–death pair with large positive and negative PCA coefficients in Fig. 3 (b).

Lower: number of children birth–death pairs of volume-optimal cycles with *n* vertices.

**Fig. 6. Traditional structural analysis for five-vertex cycles.**

a) Histogram of edge lengths in five-vertex cycles obtained from volume-optimal cycle analysis.

The width of the bin in the histogram was set to 0.01 Å and the intensity of the histogram was normalized such that the integral over the histogram became 1.

b) Histogram of angles in five-vertex cycles. The width of the bin in the histogram was set to 1°, and normalization was performed in the same manner as in (a).
c) Scatter plot of the first and second principal components obtained from PCA for the data of the length of five edges in each five-vertex cycle. The edge lengths were sorted in ascending order in each cycle. The data were preprocessed using max-min normalization prior to the PCA.
d) Scatter plot of the first and second principal components for the data of the five angles in each cycle. The preprocessing method was the same as that used for (c).
e) PD for five-vertex cycles for comparison with (c) and (d).

**Fig. 7. Analysis for children's volume-optimal cycles.**

a) Histogram of the number of vertices in the volume-optimal cycles attributed to the children's birth–death pair.
b) Distribution of children's birth–death pairs with three-, four-, five-, and six-vertex structures in the slowest-cooled models. The background is the average PD for the 20 structures used in the analysis.
c) Distribution of children's birth–death pairs with three-, four-, five-, and six-vertex structures in the fastest-cooled models. The background is the average PD for the 20 structures used in the analysis.

**Fig. 8. Ridge regression analysis using RDF.**

a) RDFs of amorphous structures generated at different cooling rates. The average of 30 samples with the same cooling rate was used in this plot.
b) Results of ridge regression between RDF and thermal conductivity with regularization parameter $\alpha = 1.0$. The same 570 structures, shown in Fig. 2, were used in this analysis. RDFs for the respective structures were obtained by binning the pair-wise distance in the range of 2.0–6.0 Å into 200 bins.
c) Distribution of the coefficient obtained from the ridge regression with $\alpha = 50.0$.
d) Results of ridge regression between RDF and thermal conductivity with regularization parameter $\alpha = 50.0$.

**Fig. 9. Relationship between volume-optimal cycles and spectral shape of RDF.**

a) Calculation results of the RDF restricted to the volume-optimal cycles that correlate with the large positive and negative first principal components in Fig. 4. The density of the original amorphous structure was used in the evaluation of the RDF. The average RDF evaluated for the 20 slowest-and fastest-cooled models was used in this plot. The insets provide examples of volume-optimal cycles in the slowest- and fastest-cooled models.

b) Calculation results of RDF restricted to five-vertex cycles in volume-optimal cycles. The inset shows an example of the cycle considered in this plot.

c) Calculation results of RDF restricted to the four-vertex cycles in the volume-optimal cycles of the children's birth–death pairs in the fastest-cooled models. The inset shows an example of the cycle considered in this plot.

**Fig. 10. Influence of five- and four-vertex cycles on thermal conductivity.**

a) Mode diffusivity in the slowest- and fastest-cooled models. The respective data for the 20 samples from each of the two models are plotted.

b) Cumulative thermal conductivity ($\kappa$) obtained in the slowest- and fastest-cooled models. The respective data for the 20 samples from each of the two models are plotted.

c) Inversion participation rate obtained in the slowest and fastest-cooled models. The respective data for the 20 samples from each of the two models are plotted.

d) Total vibrational DOS calculated for the slowest- and fastest-cooled models. The average values of the data obtained from the 20 samples from each of the two models were used for plotting.

e) Vibrational DOS projected on the atoms in the five-vertex cycles in the slowest-cooled model and the four-vertex children's cycles in the fastest-cooled model. In general, the numbers of atoms in the group of five -and four-vertex children's cycles differ. To suppress the influence of this point, the projected DOS was divided by the average number of atoms in each group after taking the average values of the 20 samples from each of the two models.

f) Histograms of the absolute value of the dynamical matrix elements correlated with the atoms in the five-vertex cycles in the both models and the four-vertex children's cycles in the fastest-cooled model.

**Fig. 11. Analysis of dynamic structure factors.**

Longitudinal and transverse components of the dynamic structure factors and the dispersion relationship, extracted by fitting the dynamic structure factor by a Lorentzian function. The dynamic structure factors were evaluated at the wave vector $|\boldsymbol{q}| = 2\pi N/L$, where $L = 4.378$ nm and $N = 1,2,\ldots,8$. The error bars in the dispersion relationship correspond to the mode lifetime.

a) Results in 4,096-atoms system generated at cooling rates of $10^{11}$ K/s.

b) Those of $10^{14}$ K/s.

Figures

Figure 1

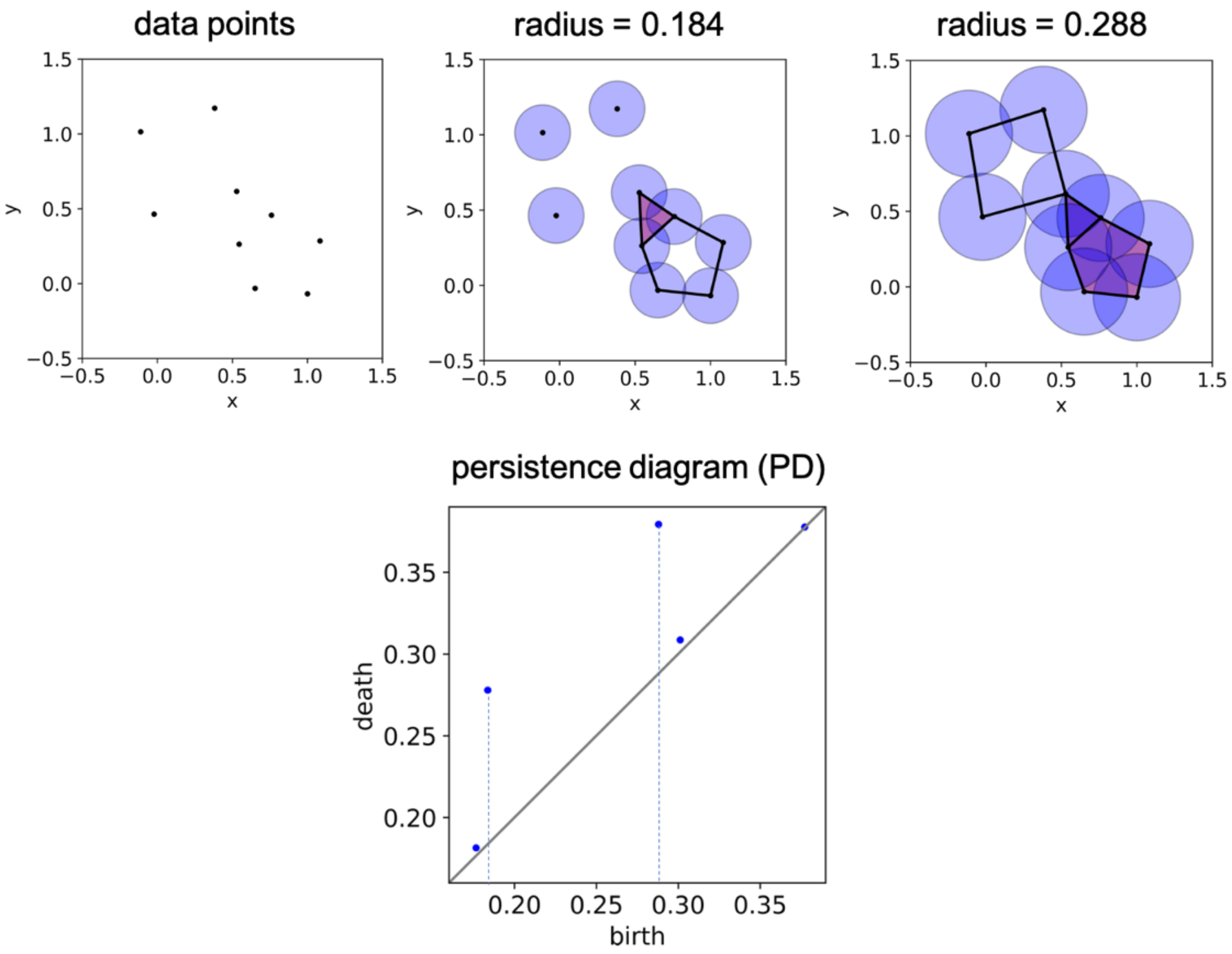

Figure 2

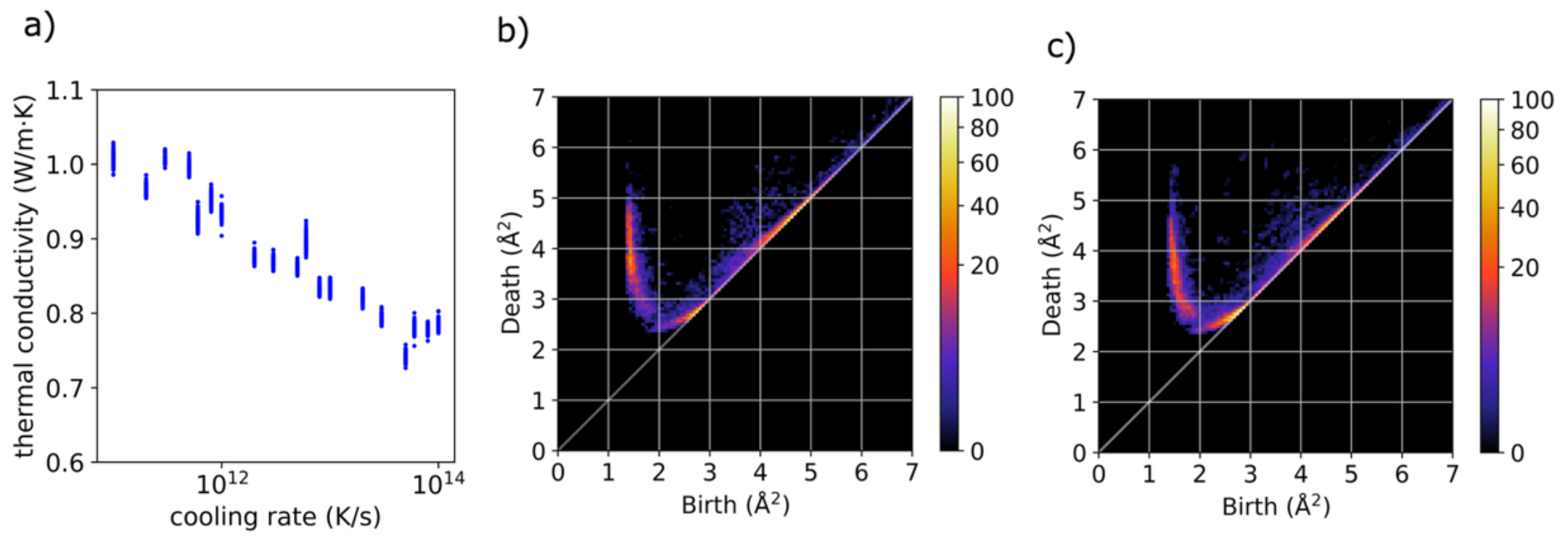

Figure 3

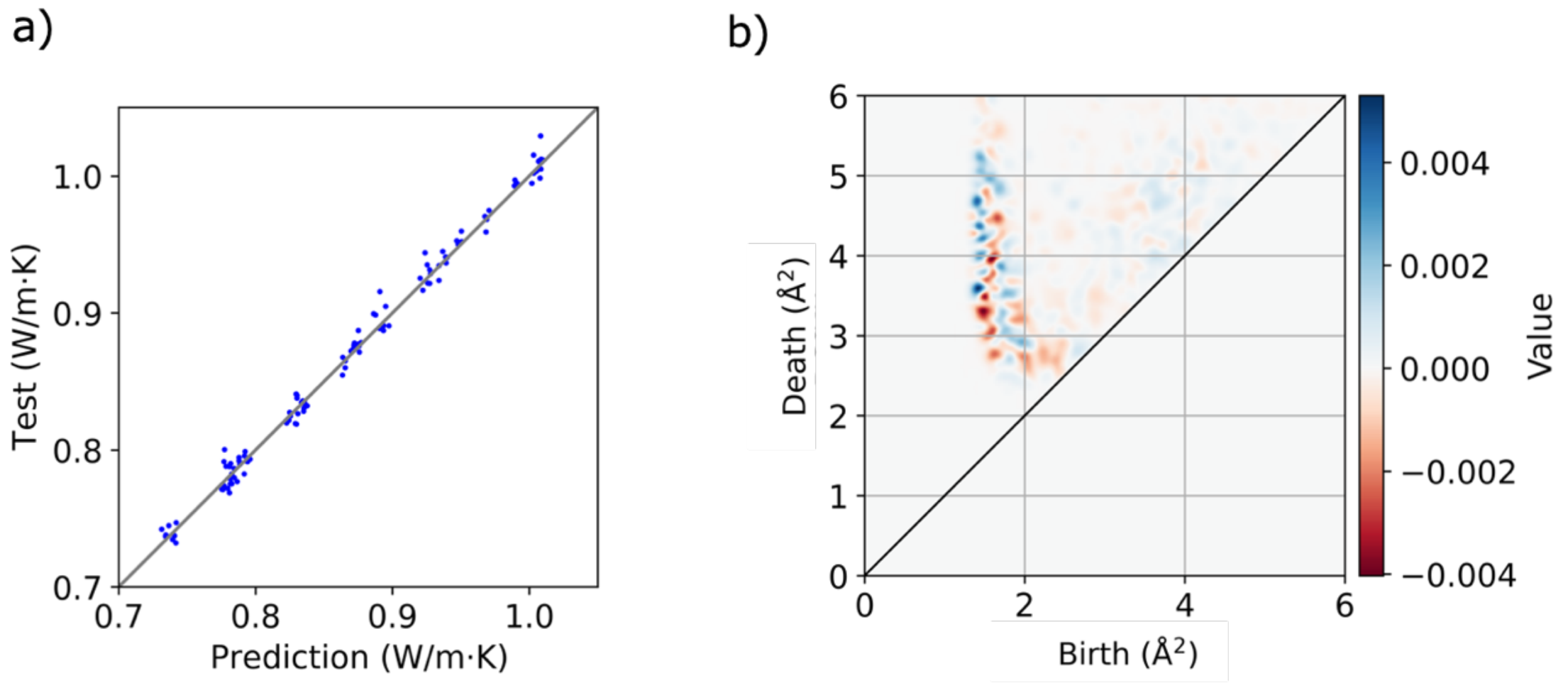

Figure 4

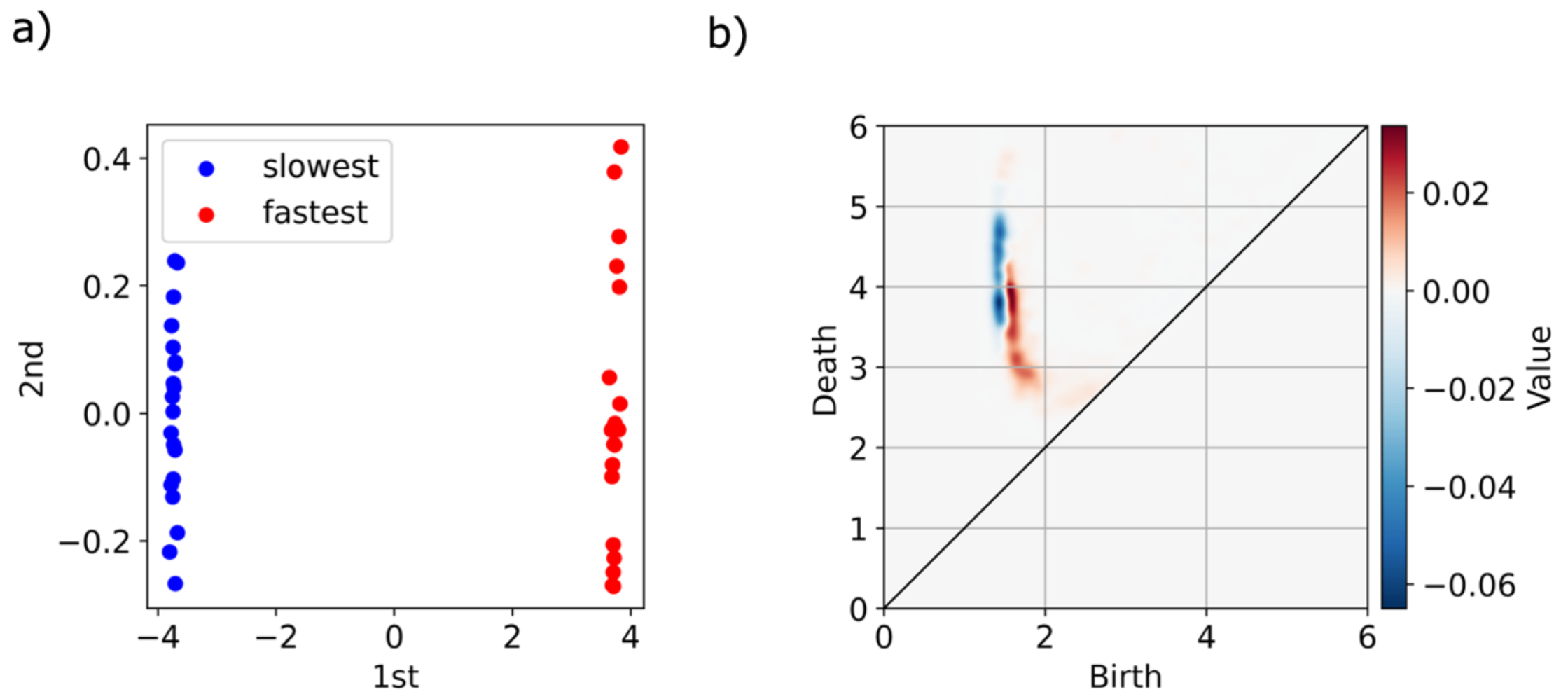

Figure 5

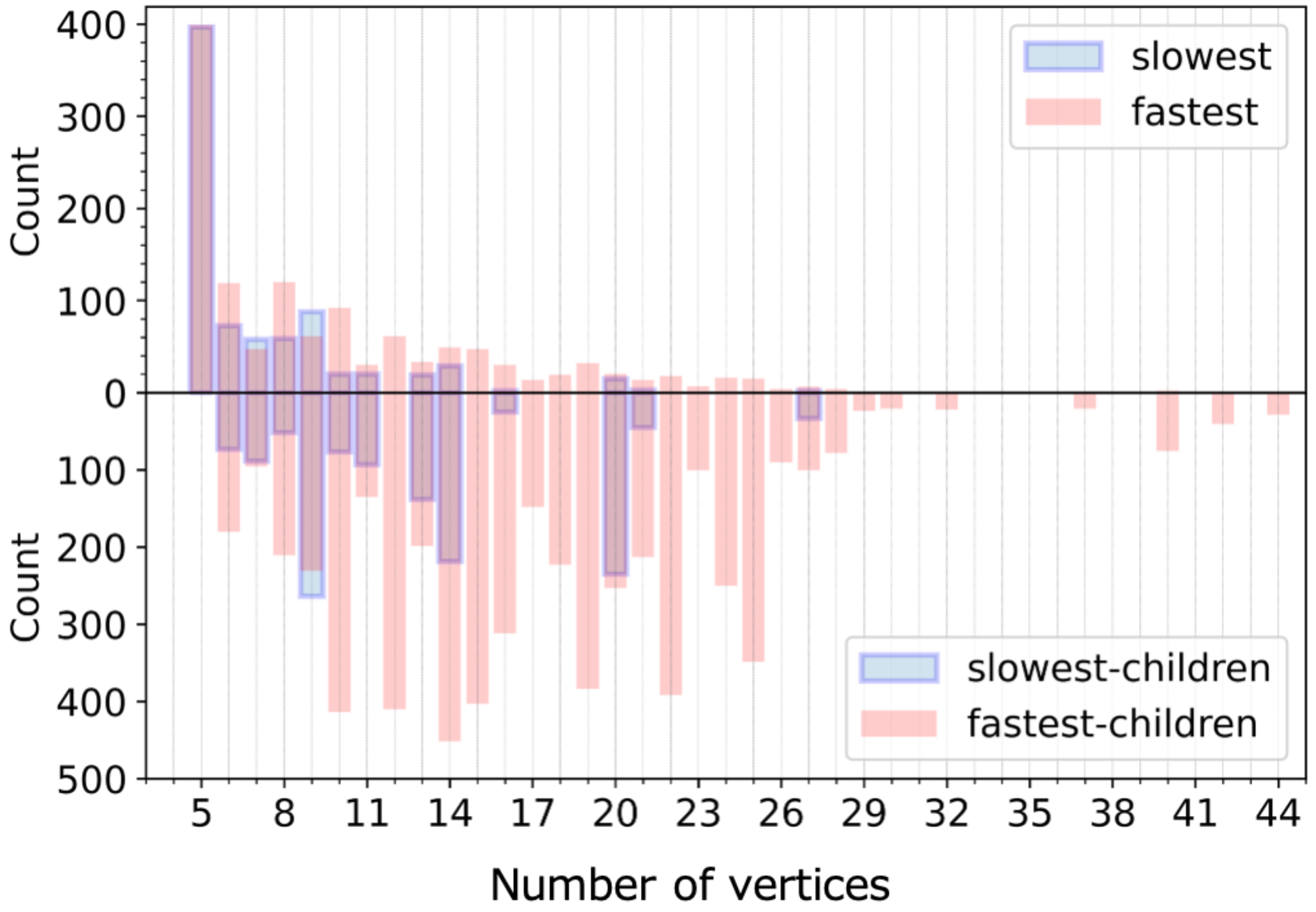

Figure 6

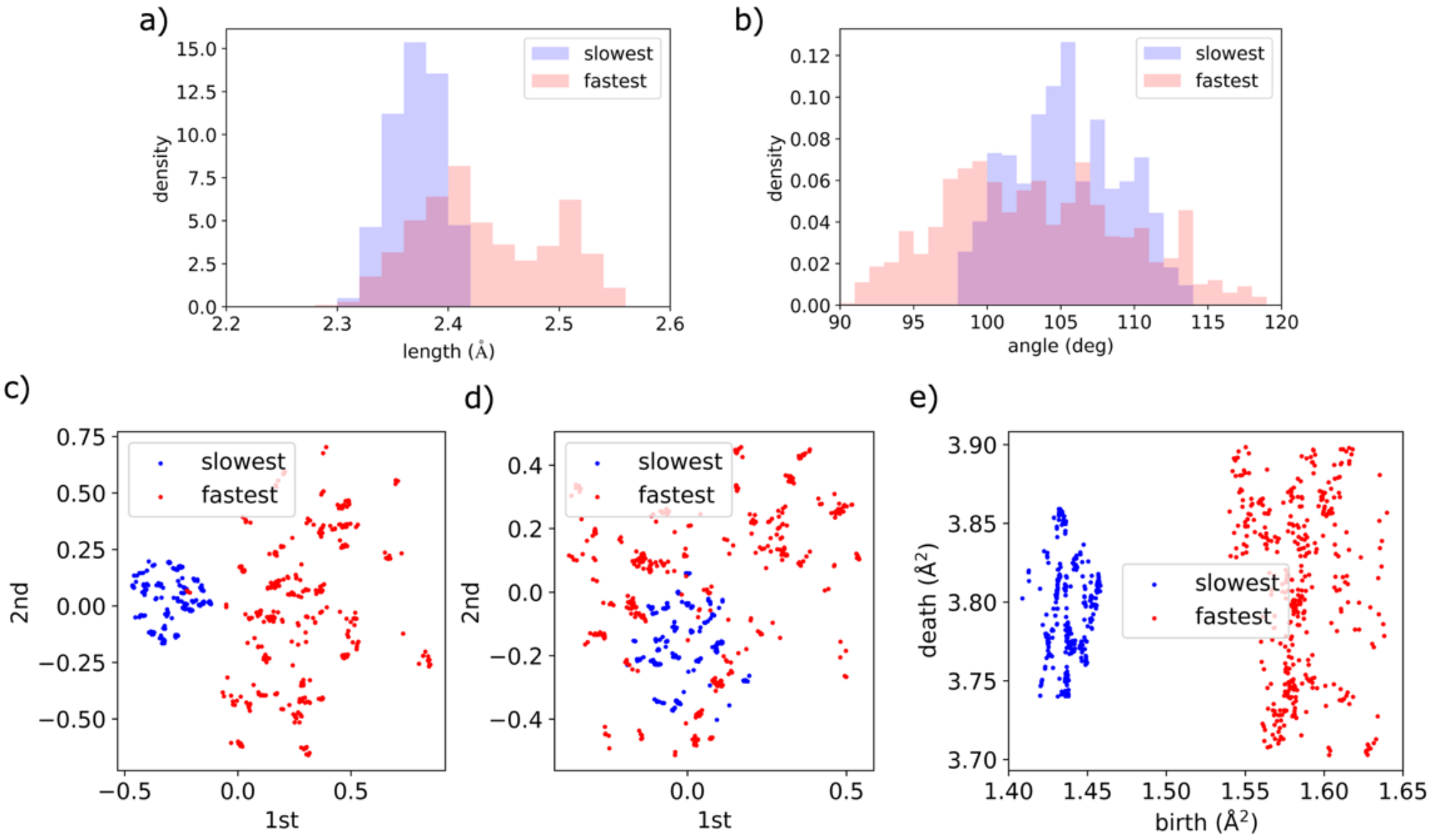

Figure 7

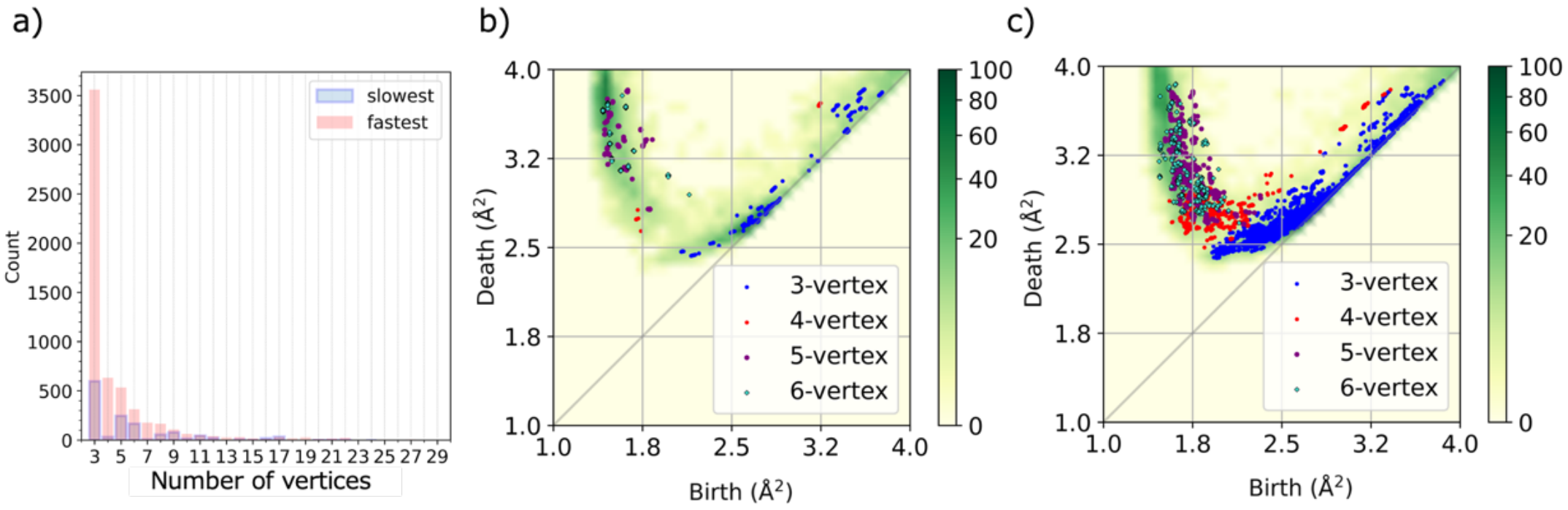

Figure 8

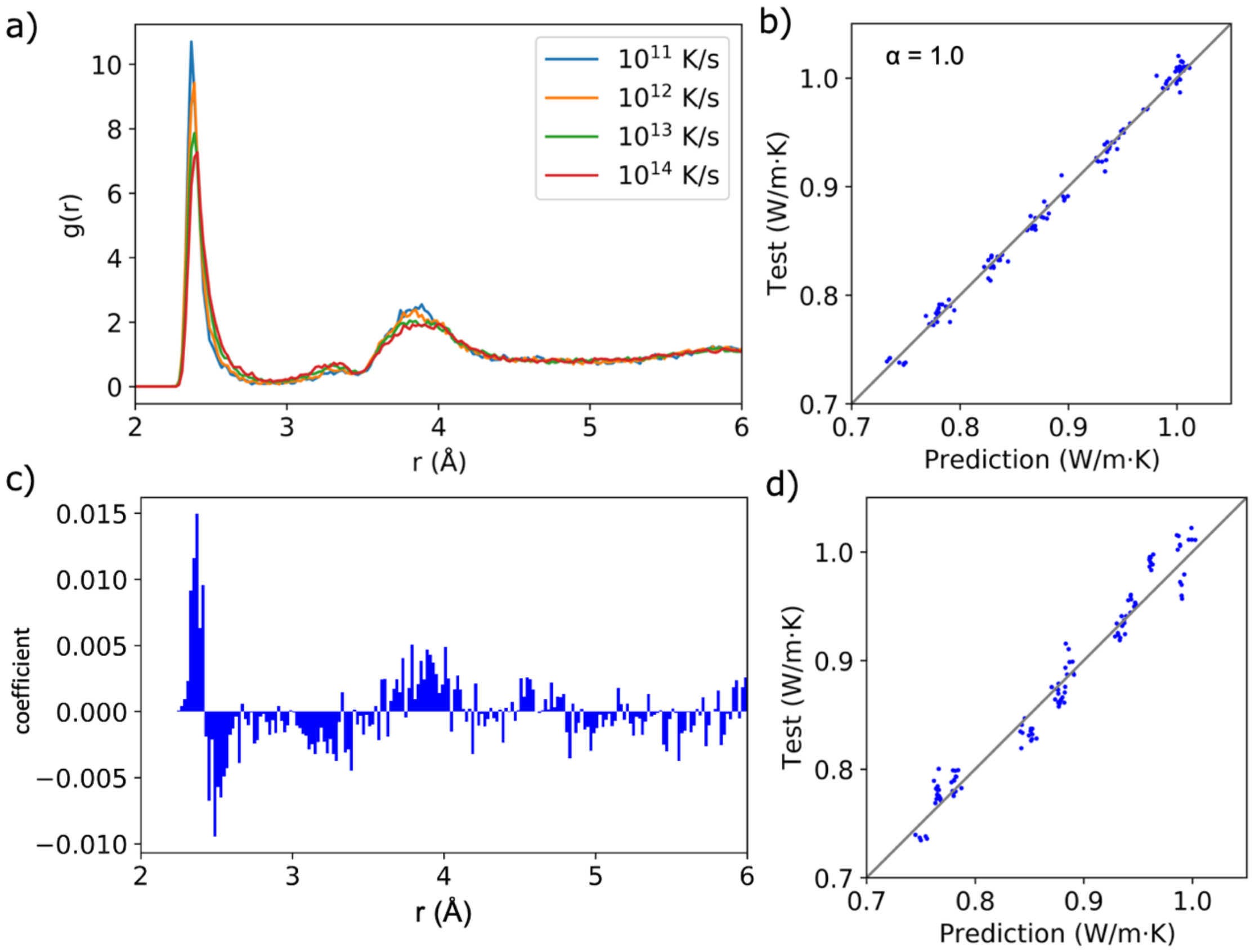

Figure 9

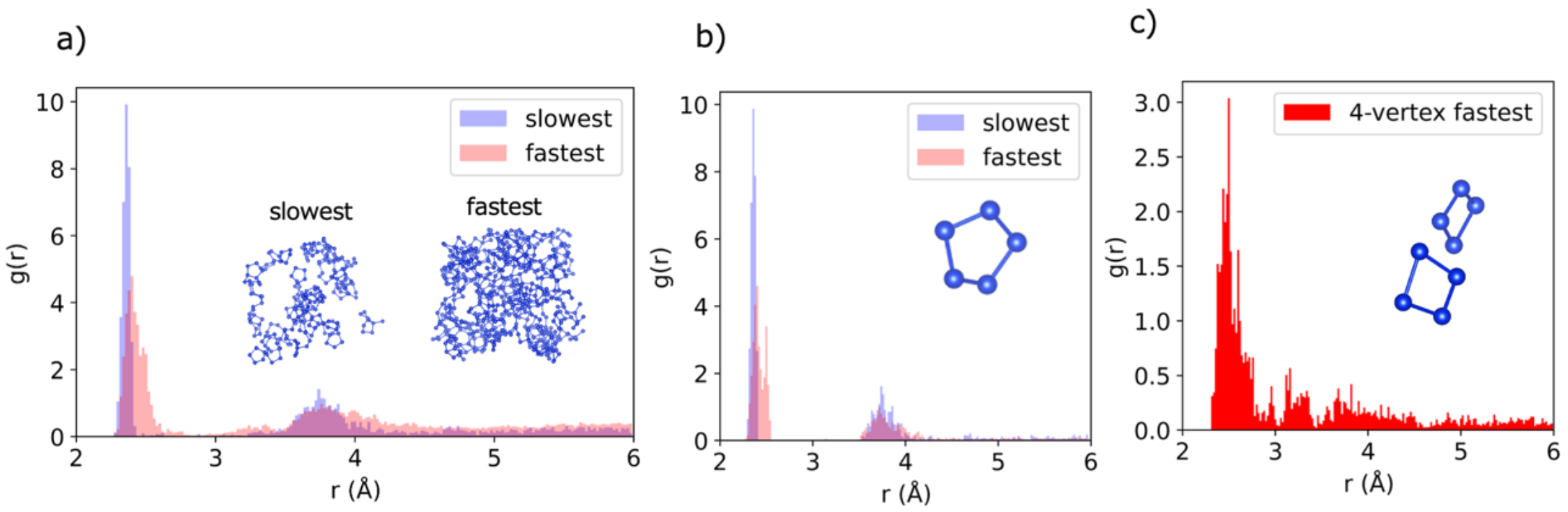

Figure 10

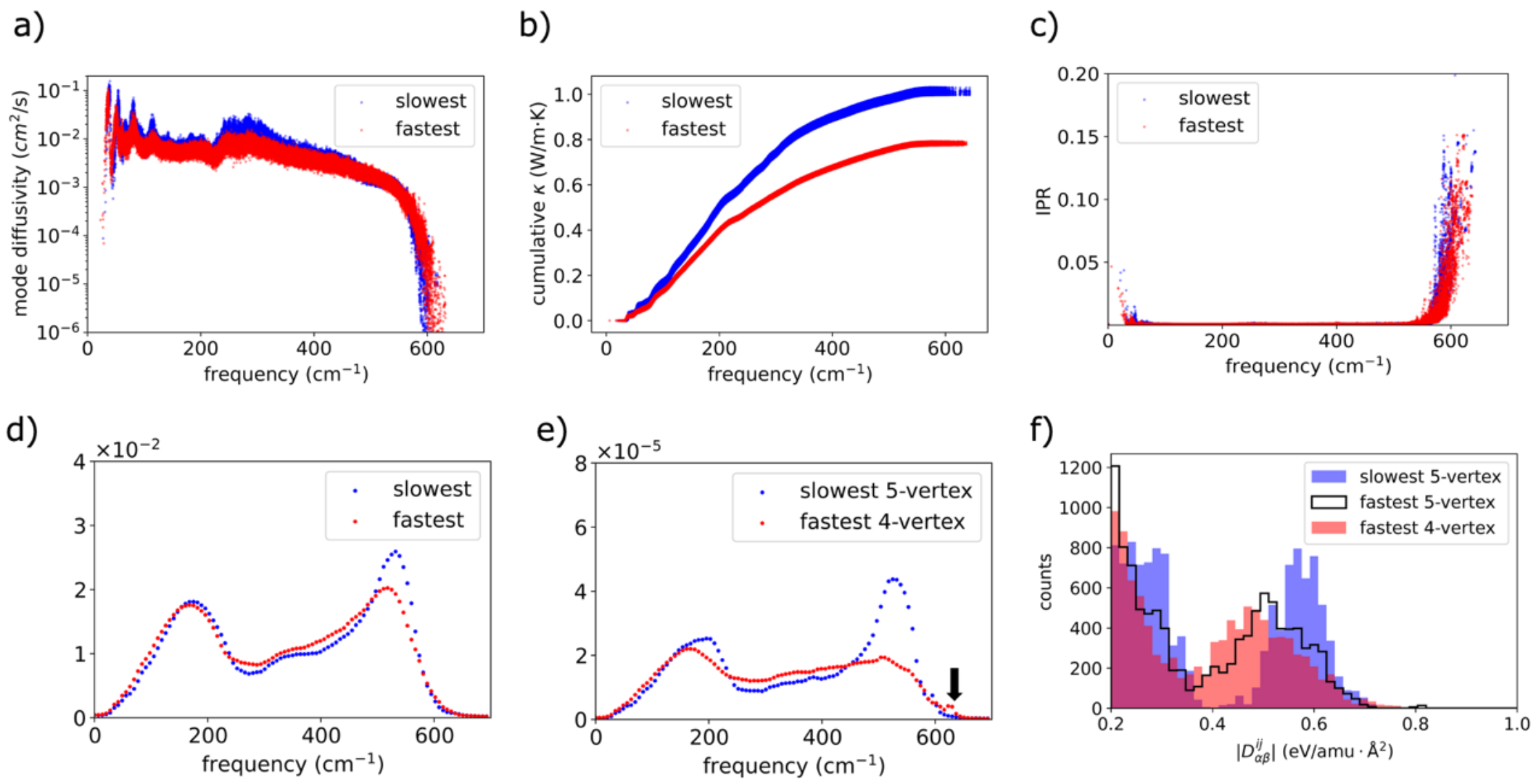

Figure 11

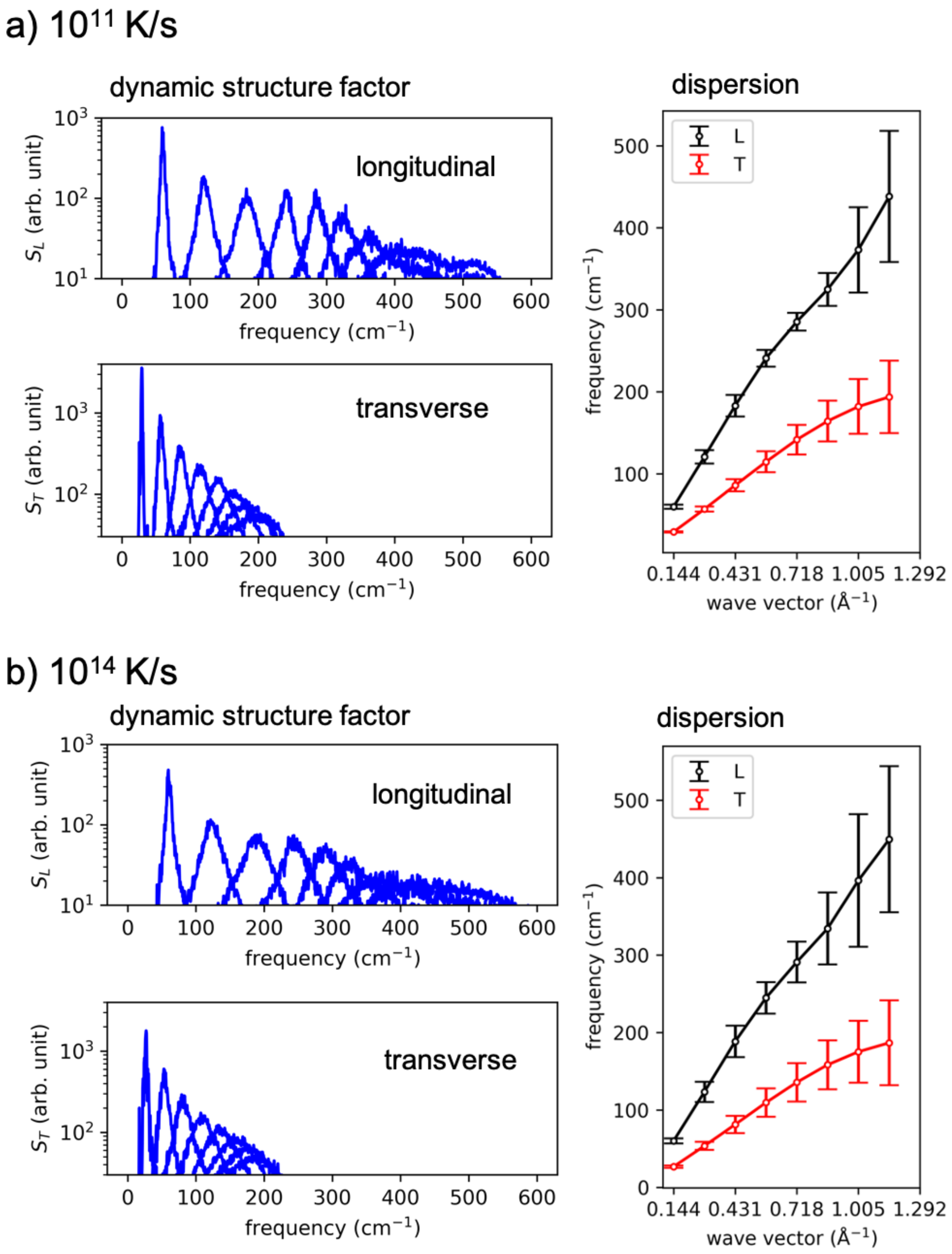